\documentstyle[aps,twocolumn]{revtex}

\begin{document} \input psfig.sty  

\title{The Effect of Large Amplitude Fluctuations
in the Ginzburg-Landau Phase Transition}

\author{G. Alvarez and H. Fort \\
Instituto de F\'{\i}sica, Facultad de Ciencias, \\ Igu\'a 4225, 11400
Montevideo, Uruguay}

\maketitle

\begin{abstract}

The lattice Ginzburg-Landau model in $d=3$ and $d=2$ is simulated
, for different values of the coherence length $\xi$ in units of the 
lattice spacing $a$, using a Monte Carlo method. The energy, 
specific heat, vortex density $v$, helicity modulus $\Gamma_\mu$ 
and mean square amplitude are measured to map
the phase diagram on the plane $T-\xi$. 
When amplitude fluctuations, controlled by the parameter
$\xi$, become large ($\xi \sim 1$) a proliferation of vortex excitations
occurs changing  the phase
transition from continuous to first order.   
\end{abstract}

PACS numbers: 74.25.Dw, 64.60.-i, 05.70.Fh

\vspace{2mm}

The Ginzburg-Landau (G-L) model, involving a complex field $\psi =
|\psi|\exp[i\theta]$, is a kind of ©multi purpose' or
©metamodel' which captures the essence of several 
interesting phenomena of condensed matter:
superfluidity, superconductivity and the melting transition.  
It turns out that in all these phase transitions vortex-like 
defects (topological singular phase configurations) play a 
central role condensing 
above the critical temperature $T_c$ and breaking the
ordered state existent below $T_c$ \cite{kle89}.
With the discovery of high-temperature superconductors  
an entire new area in condensed matter 
devoted to {\em vortex physics} has been opened \cite{nel97}
,\cite{bl-fe-ge-la-vi94}. 
A very rich phase diagram for the {\em vortex matter}
emerged both from the experimental and theoretical side.

The nature of the G-L phase transition in 
$d$=3 and $d$=2 systems still remains controversial 
theoretically. For $d$=3, in the early 70's, 
Wilson and Fisher \cite{wi-fi72}, using 
renormalization group techniques, concluded that the G-L model
belongs to the same universality class of the XY model 
("phase only" approximation), which is known to exhibit
a second-order phase transition. On the other hand, 
at $T=T_c$ amplitude fluctuations, controlled
by the coherence length in units of the lattice spacing $\xi$,
are in principle not negligible and 
they might affect the critical behavior \cite{bo-be94}.
In that sense, a recent
variational approach \cite{cu-be00} showed evidences that  
for $d=$3 the G-L transition can turn first order due to the 
interplay between phase and amplitude fluctuations when the latter are
large enough. For $d=2$ doubt remained if the first order found was  
not an artifact of the used approximation.
However, for $d=2$, we found evidences  
of a first order transition when amplitude fluctuations
are allowed to be large by a suitable choice of the G-L 
hamiltonian parameters \cite{AF99}.

In this letter, using a Monte Carlo method we show that:
1) the G-L phase transition becomes first order when
$\xi$ is greater than a value $\xi_{2 \rightarrow 1}$ 
close to 1, allowing non-negligible amplitude fluctuations
and 
2) that this change of order is connected with a sudden 
proliferation of vortices. 
We worked on square lattices of size $L$ and spacing $a$, 
we denote the lattice sites by $x$ and the lattice links 
by ($x,\mu)$ with $\mu=1,..,d$.
We used the parameterization of the Boltzmann exponent $\beta H$  
of ref. \cite{bo-be94} i.e. in 
terms of  two parameters, the coherence length
$\xi$ and a lattice temperature $T_L$,
and a dimensionless order parameter $\bar{\psi}$.
This parameterization is
connected with the ordinary Wilson parametrizarion $H_W/T$ 
\cite{ba80} by: 
$$\frac{H_W}{T} \equiv \frac{1}{T} a^d 
\sum_x [ \sum^d_{\mu=1} \frac{1}{2} \frac{(
\psi_{x+a\mu}-\psi_x)^2}{a^2} +r |\psi_x|^2 + u |\psi_x|^4 ] =$$
\begin{equation}
\frac{1}{2T_L}H \equiv \frac{1}{2T_L}
a^d \sum_x [ \sum^d_{\mu=1}  
\frac{1}{2}\frac{(\bar{\psi}_{x+a\mu}-\bar{\psi}_x)^2}{a^2} +
\frac{1}{2\xi^2}
(1-|\bar{\psi}_x|^2)^2 ],
\label{eq:H2}
\end{equation}
\begin{equation}
T_L=\frac{ u }{a^{d-2}|r|}T,
\;\;\;\;\;\;\;\;\;
\frac{\xi^2}{a^2}=\frac{1}{2 |r|};
\label{eq:connec}
\end{equation}
where $\bar{\psi}_x =\equiv \frac{\psi_x}{\psi_{\infty}}$ and
$\psi_{\infty}$ is the constant value
to what $\psi$ approaches 
infinitely deep in the interior of the superconductor.
In what follows we will omit the subscript $L$ of $T$.
In the limit of $\xi=0$ (or $u=\infty$) the radial degree of 
freedom is frozen and the model reduces to the XY 
model. The limit we are interested in is just the opposite: $\xi \sim 1$, 
where large amplitude fluctuations occur.

The calculations were 
performed using periodic boundary conditions (PBC).
As it is common practice, we have discretized
the $O(2)$ global symmetry group to a $Z(N)$ to
increase the speed of the simulation.
For $Z(60)$, we found no appreciable differences when comparing results with
previous runs carried out with the full O(2) group in relatively small 
lattices.
We used a sequence of lattice sizes 
from $L$=10 to $L$=64 for $d=2$ and from $L$=6 to $L$=16 for $d=3$. 
For the case up to $N\equiv L^d=508$ sites, we thermalized with usually
20.000-40.000 sweeps and
averaged over another 50.000-100.000 sweeps. For greater $N$   
larger runs were performed, typically 50.000 sweeps were discarded for 
equilibration and averaged over 150.000-250.000 sweeps. 
We wish to emphasize that for the case $\xi \sim 1$, we are most
interested in, 
no appreciable differences take place by taking $N>508$ (for instance
$8^3$ and $16^3$ give very similar results).
The errors for the measured observables are estimated in a standard
way by dividing measures in bins large enough to regard them as uncorrelated
samples.
The following observables were measured to map the phase diagram and 
analyze the phase transition:

i) The energy density $\varepsilon \equiv<H>/N$
and the specific heat $c_V$.
$c_V$ was computed both simply as
the energy variance per site i.e. $c_V=(<H^2>-<H>^2)/N$
and as the temperature 
derivative of $\varepsilon$.

ii) The vortex density $v$ (density of loop vortices in
$d$=3 and point vortices in $d$=2), which serves as a measure
of the phase disorder, 
defined by:
\begin{equation}
v=\frac{1}{L^2} \sum_{*p} |m_{*p}|,
\label{eq:v}
\end{equation}
where $*p$ denotes the lattice cell dual to a given lattice 
plaquette $p$ (a site in $d=2$ and a link in $d=3$) and 
the quantity $m_{*p}$ is the vortex ``charge" assigned to
$*p$ and measured over the plaquette $p$ as:
\begin{eqnarray}
m_{*p}=\frac{1}{2\pi}({[}\theta_1 - \theta_2 {]}_{2\pi} +
{[}\theta_2 - \theta_3 {]}_{2\pi} + \nonumber \\
+ {[}\theta_3 - \theta_4 {]}_{2\pi} +
{[}\theta_4 - \theta_1 {]}_{2\pi}), 
\label{eq:m}
\end{eqnarray}
where $[{\alpha}]_{2\pi}$ stands for $\alpha$ modulo 2$\pi$:
$[{\alpha}]_{2\pi} = \alpha + 2{\pi}n$, with $n$ an integer
such that $\alpha + 2{\pi}n
\in (-\pi ,\pi ]$, hence
$m_{*p}=n_{12}+n_{23}+n_{34}+n_{41}$
can take three values: 0, $\pm 1$ (the value 
$\pm 2$ has a negligible probability).

iii) The {\em helicity modulus} $\Gamma_\mu$ \cite{fi-ba-ja73}
$\Gamma_\mu$ measures the phase-stiffness along the direction
$\mu$. For a spin system with PBC 
the helicity modulus measures the cost in free energy of imposing a ``twist" 
equal to $L\delta$ in the phase between two opposite boundaries 
of the system.
$\Gamma_\mu$ is computed using the expression \cite{eb-st83}:
\begin{eqnarray}
\Gamma_\mu =
\frac{1}{N} \{ <\sum_{x} \mid \bar{\psi}_x\mid \mid \bar{\psi}_{x+\mu} 
\mid \cos (\theta_{x+\mu} - \theta_x)> - \nonumber \\
- \frac{1}{T}
<{[}\sum_{x}\mid \bar{\psi}_x \mid \mid \bar{\psi}_{x+\mu}\mid \sin
(\theta_{x+\mu} - \theta_x){]}^2>\}. 
\label{eq:helicity}
\end{eqnarray}

iv) The mean square amplitude $\rho \equiv <|\bar{\psi}|^2>$ which takes
its minimum value at the phase transition.

First, in Fig. 1 we report the phase structure of the  model in the
$(T,\xi)$ plane for $d$=3 ($\diamond$) and $d$=2 ($\bullet$). 
\begin{center}
\begin{figure}[h]
\psfig{figure=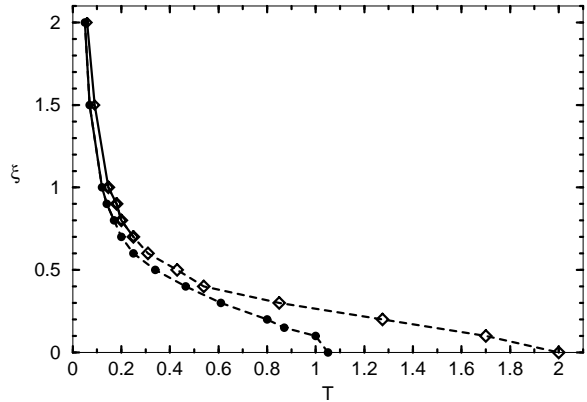,height=6cm} 
\caption{Phase diagram in the $(T,\xi)$ plane for d=2 ($\bullet$, $L=40$)  
and d=3 ($\diamond$, $L=12$).}
\end{figure}
\end{center}
For $\xi$ larger than $\xi_{2 \rightarrow 1}$ 
the order of the phase transition changes 
from continuous (dashed line) to first-order (filled line).
Specifically, $\xi_{2 \rightarrow 1}\simeq 0.8$ ($d$=2) and 
$\xi_{2 \rightarrow 1}\simeq 0.7$ ($d$=3). 
For $\xi \ge \xi_{2 \rightarrow 1}$ $\varepsilon$ exhibits
a large hysteresis phenomenon and an histogram
with a double peak structure, corresponding to the two coexisting phases, 
both features characteristic of a first-order transition.
In Fig. 2 we present energy histograms for $d$=2 for 2 values of
$\xi$:  $\xi$=0.85  $>\xi_{2 \rightarrow 1}$ and 
$\xi$=0.75 $<\xi_{2 \rightarrow 1}$,  
showing the different behavior as $L$ increases:
both peaks remain fixed as $L$ increases in the first case
while they approach each other as $L$ increases in the second case 
(besides the peaks are lower and wider). 
Furthermore, for $\xi=0.85$ the width of each
of the peaks clearly scales as 
$\sqrt(\frac{1}{L^D})=\frac{1}{L}$ 
, due to ordinary non-critical fluctuations, while for $\xi=0.75$
the peaks do not scale this way.
(Similar histograms are found for $d$=3 for 
$\xi \ge \xi_{2 \rightarrow 1}$ and $\xi < \xi_{2 \rightarrow 1}$).
\begin{center}
\begin{figure}[h]
\psfig{figure=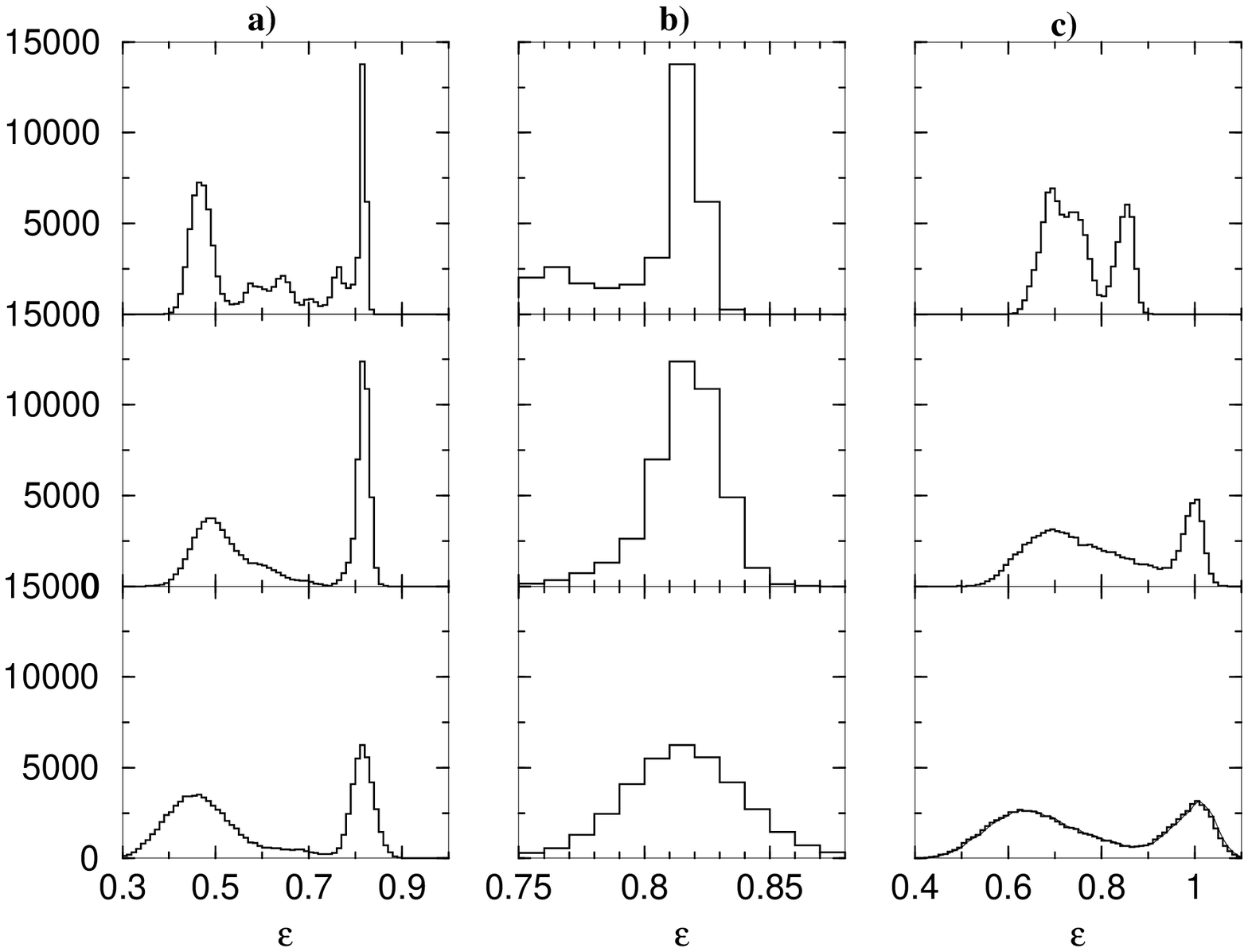,height=6cm} 
\caption{
Histograms of $\varepsilon$ in $d$=2 for
$L$=10 (below) ,$L$=20 (middle) and $L$=40 (above).
(a) $\xi$=0.85: the two peaks become sharper and their
position remain fixed as $L$ increases.
(b) Zoom of the right peak showing the scale of its width as 1/$L$.
(c) $\xi$=0.75: the width of the two peaks do not scales as 1/$L$
and they approach each other as $L$ increases.}
\end{figure}
\end{center}
Fig. 3 shows the hysteresis phenomenon found when considering 
heating and cooling runs for $d$=3 (a completely similar
hysteresis diagram is found for $d$=2 for 
$\xi \ge \xi_{2 \rightarrow 1}$).
\begin{center}
\begin{figure}[h]
\psfig{figure=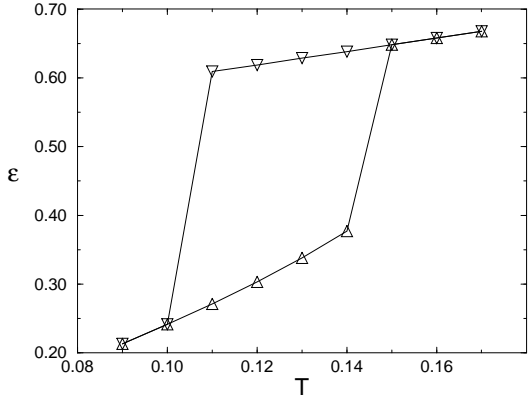,height=6cm} 
\caption{Hysteresis in $d$=3, $L$=14 for $\varepsilon$, $\xi=1$ 
(error bars are smaller than the symbol size).}
\end{figure}
\end{center}

In order to illustrate the central role played by vortex excitations
in triggering and determining the nature of the phase transition,
Fig. 4-(a) and 4-(b) 
show a plot of $v$ vs. $T$ for different values of $\xi$
respectively for $d$=2 and 
$d$=3.
For coherence lengths larger than $\xi_{2 \rightarrow 1}$  
we observe a sharp jump in the vortex density $v$ evidencing
a sudden proliferation of vortices which coincides with the
discontinuity in $\varepsilon$. As long as we
decrease $\xi$ the jump becomes more smooth and moves to higher values of 
$T_c$. In particular, for $d$=2, for $\xi\simeq 0.1$ something very
close to the Kosterlitz-Thouless (K-T) behavior is reached.
The increase in the density of vortices when amplitude fluctuations are
large is due basically to the fact that 
amplitude fluctuations decrease the energy of vortices 
enhancing vortex production.
The same happens for the XY model with modified interaction
\cite{hi84};
in fact, the shape modification of the interaction can be
connected with a core energy variation. 
\begin{center}
\begin{figure}[h]
\centering
\psfig{figure=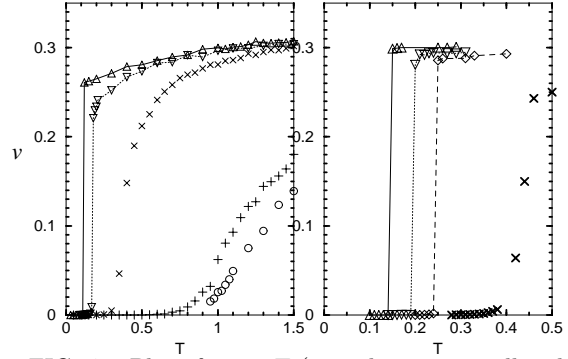,width=7.5cm}
\caption{ Plot of $v$ vs. $T$ 
(error bars are smaller than the symbol size.
Left: $d$=2, $L$=40;  $\xi$=1 ($\triangle$), 
$\xi$=0.8 ($\nabla$),
$\xi=0.5$ ($\times$), $\xi=0.1$ (+)
and XY ($\circ$).
Right: $d$=3, $L$=12; $\xi$=1 ($\triangle$), 
$\xi$=0.8 ($\nabla$), $\xi$=0.7 ($\diamond$) and
$\xi$=0.5 ($\times$) }
\end{figure}
\end{center}

\begin{figure}[h]
\psfig{figure=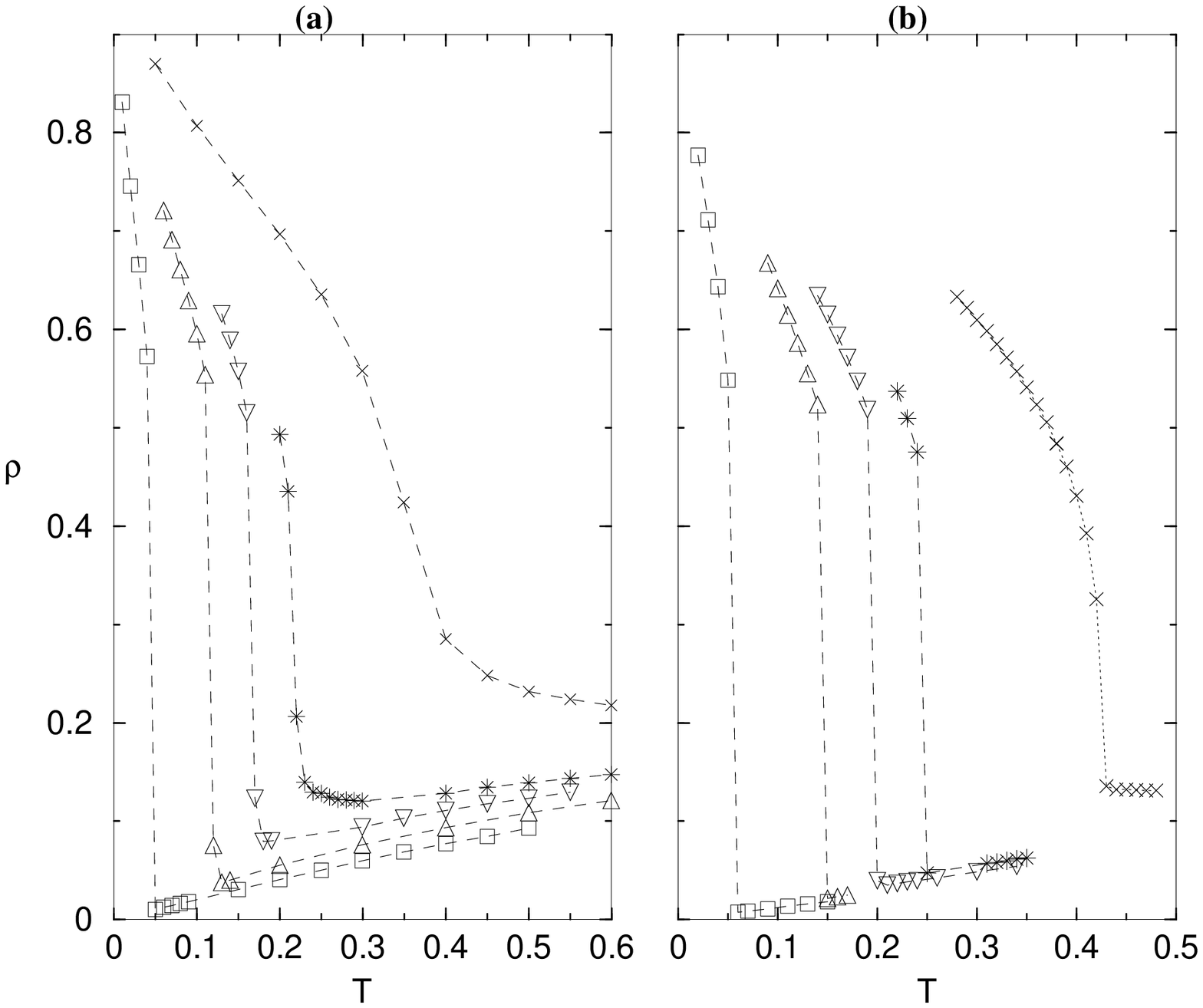,width=7.5cm}
\caption{ Plot of $<|\bar{\psi}|^2>$ vs. $T$.
(a) $d=2$: $\xi$=2 ($\Box$), $\xi$=1 ($\triangle$), 
 $\xi$=0.8 ($\nabla$), $\xi$=0.7 ($\star$) and $\xi$=0.5 ($\times$).
(b) $d=3$: $\xi$=2 ($\Box$), $\xi$=1 ($\triangle$), 
 $\xi$=0.8 ($\nabla$), $\xi$=0.7 ($\star$) and $\xi$=0.5 ($\times$).}
\end{figure}
Figures 5 shows a plot of $\rho \equiv <|\bar{\psi}|^2>$ vs. $T$ 
for different values of
$\xi$. Note that for $\xi > \xi_{2 \rightarrow 1}$ $\rho$
exhibits a sharp drop which coincides with the jump in $v$ 
and $\varepsilon$ (and the drop in $\Gamma_\mu$ not depicted). 

Therefore, the nature of the phase 
transition of the G-L model 
depends dramatically on the value of the coherence length  
$\xi$.
For $\xi > \xi_{2 \rightarrow 1}$ 
the transition to a disordered state implies latent heat, a 
a discontinuous jump of the vortex density and the subsequent
abrupt drops in $\Gamma_\mu$ and $\rho$ 
i.e. all the features of a first order transition. 
On the other hand, for $\xi \ll 1$ the G-L reduces
to the XY model (with the more subtle 
K-T phase transition for $d$=2).
For $d$=3, the value of $\xi_{2 \rightarrow 1}\simeq 0.7$
is in agreement with ref. \cite{cu-be00} in which the 
authors found the $\xi_{2 \rightarrow 1}^2 > 1/4.5$
For $d$=2, the value of $\xi_{2 \rightarrow 1}\simeq 0.8$
belongs to the region ($\xi$ >1) found 
im ref. \cite{bo-be94} such that the RG trajectories cross the
first order line of Minnhagen's
\cite{mi-wa87} generic phase diagram for the two-dimensional
Coulomb gas. 

To conclude, both for $d$=2 and $d$=3,
we offer clear evidences that the order of the
phase transition, triggered by vortices, in the G-L model 
depends on the value of $\xi$ which controlls the size of
amplitude fluctuations.
Whether or not such mechanism based on the interplay between
amplitude and phase fluctuations takes place in real systems like
high $T_c$ superconductors or in explaining the change in order of
the melting transition is something which deserves a more
careful analysis.

\vspace{1mm}

Work supported in part by 
PEDECIBA.  

We are indebted with  H. Beck and P. Curty
for valuable discussions.

\end{document}